\documentclass[10pt,letterpaper,twocolumn,english,prl,aps,showpacs,floatfix]{revtex4}
\usepackage[]{fontenc}
\usepackage[latin1]{inputenc}
\usepackage{amsmath}
\usepackage{amssymb}

\makeatletter

\def\1{1\negthickspace{\rm I}}

\usepackage{babel}
\makeatother
\begin{document}

\title{Analytic Relations between Localizable Entanglement and String Correlations\\
in Spin Systems}

\author{L. Campos Venuti}

\affiliation{Dipartimento di Fisica, V.le C. Berti-Pichat 6/2, I-40127 Bologna,
Italy}

\author{M. Roncaglia}

\affiliation{Dipartimento di Fisica, V.le C. Berti-Pichat 6/2, I-40127 Bologna,
Italy}

\date{\today}

\begin{abstract}
We study the relation between the recently defined localizable entanglement
and generalized correlations in quantum spin systems. Differently
from the current belief, the localizable entanglement is always given
by the average of a generalized string. Using symmetry arguments we
show that in most spin-1/2 and spin-1 systems the localizable entanglement
reduces to the spin-spin or string correlations, respectively. We
prove that a general class of spin-1 systems, which includes the Heisenberg
model, can be used as perfect quantum channel. These conclusions are
obtained in analytic form and confirm some results found previously
on numerical grounds.
\end{abstract}


\maketitle
Entanglement is one of the most specific features of Quantum Mechanics
(QM), first recognized by Schr\"{o}dinger \cite{schroedinger35},
that has been realized in recent times to open the way to some of
the most important applications of QM like teleportation, dense coding,
quantum cryptography and computation \cite{heissbook}. Since quantum
information is based on qbits and, more generally on finite-level
quantum systems, spin chains emerge as natural candidates for quantum
devices. Moreover, it has also been suggested recently that the quantum
information perspective could give a better understanding of the structure
of the ground state and of possible quantum phase transitions in Condensed
Matter systems \cite{osborne02,vidal03,fan04}. In this context Verstraete
\emph{et al.}~have introduced the notion of localizable entanglement
(LE) \cite{verstraete04a}, which is a very attracting new concept
for its possible applications to the study of quantum phase transition.
In addition, the LE has a precise physical meaning to the extent that
it can measure the performance of certain kinds of quantum repeaters
\cite{briegel98,zeilinger-danube}. 

The LE of an interacting multipartite system is the maximum amount
of entanglement that can be localized on two particles by performing
\emph{local} measurements on the remaining (assisting) particles.
Here we stick to the class of projective Von Neumann measurements.
The LE on the pair $i,j$ of an $N+2$ particle state $|\psi\rangle$
is defined as \begin{equation}
L_{i,j}\left(|\psi\rangle\right)=\max_{\left\{ |s\rangle\right\} }\sum_{s}p_{s}E\left(|\varphi_{s}\rangle\right).\label{eq:LE}\end{equation}
 The maximum is taken over all local measurement bases $\left\{ |s\rangle\right\} $
of the $N$ assisting particles, $|\varphi_{s}\rangle$ is the normalized
two-qbit state at sites $i$ and $j$ obtained after performing the
measurement $|s\rangle$ with outcome probability $p_{s}$. As a measure
of entanglement $E\left(|\varphi_{s}\rangle\right)$ for two-qbit
states $|\varphi_{s}\rangle$ we use here the concurrence, first defined
in \cite{wootters98}. Finding the optimal basis which satisfies Eq.~(\ref{eq:LE})
is in general a formidable task. However in \cite{verstraete04a}
upper and lower bounds have been derived, the lower bound being given
by the maximal connected correlation between sites $i$ and $j$,
while the upper bound is the concurrence of assistance \cite{divincenzo98}. 

Interestingly, it has been observed numerically that in most spin-1/2
systems the LE equals the lower bound \cite{popp04}. Similar numerical
studies have shown \cite{verstraete04b,popp04} that the LE assumes
its maximal value 1 for the ground state of the spin-1 Heisenberg
model with spin 1/2 at the endpoints. The same result has been proved
exactly \cite{verstraete04b} for the Affleck-Kennedy-Lieb-Tasaki
(AKLT) model \cite{AKLT}. Instead the $\phi$-deformed AKLT model
\cite{verstraete04b} exhibits an exponentially decaying LE with a
finite entanglement length $\xi_{E}$, which is defined in analogy
with the correlation length. For the latter model the LE has been
considered an essential indicator of the transition taking place at
$\phi=0$, since here $\xi_{E}$ diverges while the standard correlation
length is smooth. To rule out the possibility that the LE can be ascribed
to the string order correlation, the authors of Ref.~\cite{popp04}
have given an example of a state for which $\xi_{E}$ diverges even
in absence of string order.

In this Letter, it is shown that the LE is given by a generalized
string whose form depends on the symmetry properties of the state.
In particular the following analytical results are given: i) in spin
1/2 systems we give arguments that show the tightness of the lower
bound ii) for a large class of spin 1 systems, that include the Heisenberg
model, the LE is proved to be 1, i.e.~maximal; iii) in the $\phi$-deformed
AKLT model we show that the LE is connected with a string order correlation.

Throughout the text we consider a system of $N$ assisting particles
labeled from $1$ to $N$ and two qbits on which we localize the entanglement
that, without loss of generality, we place at the endpoints $0,\, N+1$.

Once an optimal basis $\left\{ |s\rangle=|s_{1}\ldots s_{N}\rangle\right\} $
is found, the LE of an $N+2$ particle state $|\psi\rangle$ is given
by\[
L\left(|\psi\rangle\right)=\sum_{s}\left|P\left(|\phi_{s}\rangle\right)\right|,\,\,\,|\phi_{s}\rangle\equiv\langle s|\psi\rangle,\]
where the preconcurrence $P$ of an unnormalized two-qbit state $|\phi\rangle$,
is defined by $P\left(|\phi\rangle\right)=\langle\phi^{\ast}|\sigma^{y}\otimes\sigma^{y}|\phi\rangle,$
and the complex conjugate is taken in the standard basis. It is then
clear that the LE can always be written as an expectation value\begin{equation}
L\left(|\psi\rangle\right)=\langle\psi|\sigma_{0}^{y}G^{\left(s\right)}\sigma_{N+1}^{y}|\psi^{\ast}\rangle\end{equation}
 where the operator $G^{\left(s\right)}$ is given by\begin{equation}
G^{\left(s\right)}=\sum_{s}|s\rangle\langle s^{\ast}|\,\textrm{sign}\left(P\left(|\phi_{s}\rangle\right)\right),\label{eq:G-op}\end{equation}
 and the sign function is $\textrm{sign}\left(z\right)=z/\left|z\right|$
for nonzero $z$, $\textrm{sign}\left(0\right)=0$. As it is evident
from the definition, the $G^{\left(s\right)}$ operator depends on
the optimal basis as well as on $|\psi\rangle$. However, in most
cases $G^{\left(s\right)}$ takes a manageable form that depends on
the symmetry property of the state. On top of that, if the signs in
Eq.~(\ref{eq:G-op}) factorize into local terms, then $G^{\left(s\right)}$
becomes a string $G^{\left(s\right)}=G_{1}G_{2}\cdots G_{N}$. 

In order to find the optimal basis $\left\{ |s\rangle\right\} $ we
take an infinitesimal variation over all possible local unitary transformations
obtaining the following set of extremal equations:\begin{align}
\sum_{s}\textrm{Im}\left[\Gamma_{i}^{\mu}\left(s\right)\textrm{sign}\left(P\left(|\phi_{s}\rangle\right)\right)\right] & =0\label{eq:extremalA}\\
\Gamma_{i}^{\mu}\left(s\right) & =0,\,\,\,\textrm{for}\,\, P\left(|\phi_{s}\rangle\right)=0,\label{eq:extremalB}\end{align}
 where we have defined\[
\Gamma_{i}^{\mu}\left(s\right)=\langle\psi|\sigma_{0}^{y}\left[\tau_{i}^{\mu}|s\rangle\langle s^{\ast}|+|s\rangle\langle s^{\ast}|\left(\tau_{i}^{\mu}\right)^{\ast}\right]\sigma_{N+1}^{y}|\psi^{\ast}\rangle,\]
 and the $\tau^{\mu}$'s are the generators of the unitary single-site
group $SU\left(D\right)$ ($D$ is the dimension of the single site
Hilbert space), $\mu=1,\ldots,D^{2}-1$ and $i=1,\ldots,N$. 

In what follows we shall discuss separately spin 1/2, spin 1 and matrix
product states. 

\emph{Spin 1/2.} In this case the $\tau^{\mu}$'s are the Pauli matrices.
On the basis of Eqs. (\ref{eq:extremalA}) and (\ref{eq:extremalB}),
it is straightforward to show that if $|\psi\rangle$ is real and
invariant under $\pi$ rotations about the $\alpha$-axis, $P^{\alpha}=\prod_{i=0}^{N+1}\sigma^{\alpha}$,
$\alpha=x,y,z$ then an extremal is provided by the $\sigma^{\alpha}$
basis. This result can readily be applied, for example, to the Ising
model in transverse field and to the XXZ Heisenberg model.

Let us consider first the Ising model in transverse field: \begin{equation}
H=-\lambda\sum_{i}\sigma_{i}^{x}\sigma_{i+1}^{x}-\sum_{i}\sigma_{i}^{z}.\label{eq:ising}\end{equation}
 For finite $N$ the ground state $|\psi\rangle$ is unique for any
$\lambda$ and belongs to the $P^{z}=1$ sector. For $\left|\lambda\right|>1$
the gap between the lowest state in the sector $P^{z}=-1$ and the
ground state vanishes exponentially with $N$ and hence becomes degenerate
with it in the thermodynamic limit. As the basis of $\sigma^{z}$
is extremal, we calculate now $G^{\left(z\right)}$. For $\lambda>0$
a Marshall's sign theorem ensures that, in the $\sigma^{z}$ basis,
all the coefficients of $|\psi\rangle$ in the same parity sector
are non-zero and have the same sign. Then the preconcurrence $P\left(|\phi_{s}\rangle\right)$
is positive (negative) according to whether the parity of $|s\rangle$
is $-1$ ($1$). The case $\lambda<0$ is reduced to the previous
one by transforming the Hamiltonian (\ref{eq:ising}) with a $\pi$
rotation about the $z$ axis on the even sites. Collecting results,
for any non-zero $\lambda$, we get \begin{equation}
G^{\left(z\right)}=-\textrm{sign}\left(\lambda\right)^{N+1}\prod_{i=1}^{N}\sigma_{i}^{z}.\end{equation}
 This implies at once that the LE calculated in this basis is $L^{\left(z\right)}\left(|\psi\rangle\right)=\textrm{sign}\left(\lambda\right)^{N+1}\langle\psi|\sigma_{0}^{x}\sigma_{N+1}^{x}|\psi\rangle$,
i.e.~the maximal classical correlation, that is the lower bound.
For $\left|\lambda\right|>1$, in the thermodynamic limit we must
be cautioned that the LE depends on the particular combination of
the twofold degenerate ground states one chooses.

Let us discuss now the XXZ model\begin{equation}
H=\sum_{i}\left(\sigma_{i}^{x}\sigma_{i+1}^{x}+\sigma_{i}^{y}\sigma_{i+1}^{y}+\Delta\sigma_{i}^{z}\sigma_{i+1}^{z}\right).\end{equation}
The Hamiltonian commutes with all $P^{x,y,z}$ discrete rotations,
meaning that $\sigma^{x}$'s , $\sigma^{y}$'s and $\sigma^{z}$'s
are all extremal bases. Which one yields the highest LE must be assessed
by explicit calculation. Exploiting the Marshall's sign property we
get $G^{\left(z\right)}=-\left(-1\right)^{N}\mathcal{P}_{1,N}^{0}$,
where $\mathcal{P}_{1,N}^{0}$ is the projector onto the subspace
of zero magnetization for the assisting spins. This means that the
LE assumes the form \begin{equation}
L^{\left(z\right)}\left(|\psi\rangle\right)=-\left(-1\right)^{N}\langle\psi|\sigma_{0}^{y}\sigma_{N+1}^{y}|\psi\rangle.\end{equation}
 To evaluate $G^{\left(x\right)}$ we need the Marshall's sign property
in the $\sigma^{x}$'s basis. This can be achieved with a local unitary
transformation which depends on the value of $\Delta$. For $\Delta>1$
the required rotation is $\prod_{j}\exp\left(i\pi\sigma_{2j}^{x}/2\right)$
which acts only on one sublattice. Instead for $-1<\Delta<1$ we must
further apply a uniform rotation of $\pi/2$, $\prod_{j}\exp\left(i\pi\sigma_{j}^{x}/4\right)$.
For $N$ is even, the result is \begin{equation}
G^{\left(x\right)}=-\left(-1\right)^{N/2}\prod_{j=1}^{N}\sigma_{j}^{x},\,\, L^{\left(x\right)}=-\langle\psi|\sigma_{0}^{z}\sigma_{N+1}^{z}|\psi\rangle,\end{equation}
 for $\Delta>1$, and \begin{equation}
G^{\left(x\right)}=-\1,\, L^{\left(x\right)}=-\langle\psi|\sigma_{0}^{y}\sigma_{N+1}^{y}|\psi\rangle,\end{equation}
 for $\left|\Delta\right|<1$. Clearly $L^{\left(y\right)}=L^{\left(x\right)}$
for symmetry.

In the critical regime, $-1<\Delta\leq1$, it is well known that the
transverse correlations dominate at least asymptotically, while the
$z-z$ correlations are the highest in the antiferromagnetic phase
$\Delta>1$. The conclusion is that for the ground state of the XXZ
model the $\sigma^{x}$'s basis always yields the highest LE. As for
the Ising model, the lower bound is attained, in agreement with the
numerical simulations in Ref.~\cite{popp04}, indicating that our
local maximum is indeed a global one. It is remarkable that on the
each side of the isotropic point $\Delta=1$ the LE at finite size
is given by different classical correlators. In particular it has
been observed in finite size systems that the LE between nearest-neighbour
sites displays a singularity \cite{popp04}. In our opinion, this
fact is due to the different transformations required by the Marshall's
theorem and cannot be attributed to the Berezinskii-Kosterlitz-Thouless
(BKT) transition occurring at the same point. 

Work is in progress to extend the method outlined here in presence
of a magnetic field that breaks the $P^{x}$ symmetry. Nonetheless,
the numerical computation of \cite{popp04} anticipates that for some
values of the field, the LE is strictly greater than its lower bound.

\emph{Spin 1.} Here we consider spin one assisting particles $S_{i}^{\alpha},\, i=1,\ldots,N$,
and two qbits at the endpoints. The extremal equations are still given
by Eqs.~(\ref{eq:extremalA}) and (\ref{eq:extremalB}) where the
$\tau^{\mu}$'s are now the generators of $SU\left(3\right)$, i.e.~the
eight Gell-Mann matrices . When both the state $|\psi\rangle$ and
the basis $|s\rangle$ are real, it is enough to consider in the extremal
Eq.~(\ref{eq:extremalA}) only the restricted set of purely imaginary
generators of $SU\left(3\right)$: $\left(\tau^{\rho}\right)_{\mu,\nu}=-i\epsilon^{\mu\nu\rho}$,
where $\epsilon^{\mu\nu\rho}$ is the Levi-Civita symbol.

Given the richer structure of $SU\left(3\right)$ with respect to
$SU\left(2\right)$, the invariance of the state through a rotation
of $\pi$ around $\alpha$-axis, $\Pi^{\alpha}=e^{i\pi S_{\textrm{tot}}^{\alpha}}$,
is not sufficient to prove that the basis of $S^{\alpha}$ is extremal,
at variance with the spin 1/2 case. Instead, if the state $|\psi\rangle$
is symmetric with respect to rotations of $\pi$ about \emph{two}
(and hence any) axes, using the following (anti)commutation rules
\[
\left\{ \tau^{1,3},e^{i\pi S^{z}}\right\} =\left\{ \tau^{2},e^{i\pi S^{x,y}}\right\} =\left[\tau^{2},e^{i\pi S^{z}}\right]=0,\]
 Eq.~(\ref{eq:extremalA}) is satisfied provided that the measurement
basis fulfills\begin{equation}
\left[|s\rangle\langle s|,e^{i\pi S^{x,y,z}}\right]=0.\label{eq:comm}\end{equation}
 The solution of (\ref{eq:comm}) is (apart from unimportant phase
factors) $\left\{ |\tilde{s}\rangle\right\} =\left\{ |0\rangle,|\pm\rangle\right\} \equiv\left\{ |0\rangle,\left(|+1\rangle\pm|-1\rangle\right)/\sqrt{2}\right\} $,
since the matrices $e^{i\pi S^{x,y,z}}$ are all diagonal in this
basis. 

Surprisingly, without making any further assumption on the state $|\psi\rangle$
it is possible to calculate exactly the LE in the basis $\left\{ |\tilde{s}\rangle\right\} $,
finding that it reaches its maximal value, i.e.~1. The proof goes
as follows. We write the state $|\psi\rangle$ as \begin{equation}
|\psi\rangle=\sum_{s,\sigma}C_{s,\sigma}|s_{1}s_{2}\cdots s_{N}\rangle|\sigma_{0}\sigma_{N+1}\rangle,\label{eq:psi}\end{equation}
where $|s_{i}\rangle\in\left\{ |0\rangle,|\pm\rangle\right\} $ and
$|\sigma\rangle$ is the $\sigma^{z}$'s basis for the two qbits.
Due to the invariance of the state with respect to rotations of $\pi$
about any axis, one has $e^{i\pi S_{\textrm{tot}}^{\alpha}}\psi=p_{\alpha}\psi,\,\, p_{\alpha}=\pm1$
and the coefficients in (\ref{eq:psi}) satisfy $C_{s,\sigma}=-p_{z}\sigma_{0}\sigma_{N+1}\left(-1\right)^{n_{+}\left(s\right)+n_{-}\left(s\right)}C_{s,\sigma}$
and $C_{s,\sigma}=-p_{x}\left(-1\right)^{n_{-}\left(s\right)+n_{0}\left(s\right)}C_{s,\bar{\sigma}}$.
Here $n_{\gamma}\left(s\right)$ is the number of times $s_{i}=\gamma$
in the state $|s_{1}s_{2}\cdots s_{N}\rangle$ and $|\bar{\sigma}\rangle$is
obtained by spin flipping the state $| \sigma\rangle$. The LE can
be written as \begin{multline*}
L^{\left(\tilde{s}\right)}\left(\psi\right)=2\sum_{s}\left|C_{s,\uparrow\uparrow}C_{s,\downarrow\downarrow}\right|+2\sum_{s}\left|C_{s,\uparrow\downarrow}C_{s,\downarrow\uparrow}\right|=\\
-2p_{x}\sum_{s}\left(-1\right)^{n_{+}\left(s\right)+n_{0}\left(s\right)}\left(C_{s,\uparrow\uparrow}C_{s,\downarrow\downarrow}+C_{s,\uparrow\downarrow}C_{s,\downarrow\uparrow}\right).\end{multline*}
 Now we use $\left|C_{s,\sigma}C_{s,\bar{\sigma}}\right|=-\left(-1\right)^{n_{+}\left(s\right)+n_{0}\left(s\right)}C_{s,\sigma}C_{s,\bar{\sigma}}$
and $\left(-1\right)^{n_{+}\left(s\right)+n_{0}\left(s\right)}=\langle s|\prod_{i=1}^{N}e^{i\pi S_{i}^{x}}|s\rangle$.
Finally, one obtains \begin{equation}
L^{\left(\tilde{s}\right)}\left(\psi\right)=p_{\alpha}\langle\psi|e^{i\pi S_{\textrm{tot}}^{\alpha}}|\psi\rangle=1,\quad\alpha=x,y,z.\label{eq:LE1}\end{equation}
 This implies that Eq.~(\ref{eq:extremalB}) is automatically satisfied.
In this derivation it has not been necessary to compute the $G$ operator
which anyway is given by $G^{\left(\tilde{s}\right)}=-p_{y}\mathcal{P}\prod_{i=1}^{N}e^{i\pi S_{i}^{y}}\mathcal{P}$,
where $\mathcal{P}$ projects out the states for which the preconcurrence
is zero. Along the same reasoning it can be shown that Eq.~(\ref{eq:LE1})
remains valid for general complex $|\psi\rangle$.

The finding of Eq.~(\ref{eq:LE1}) demonstrates that, in a spin 1
state the entanglement is fully localizable on the endpoints, with
the only requirement of invariance with respect to $\pi$ rotation
around any axis. This property is shared by ground states of a large
class of systems which can be regarded as perfect quantum channels. 

An example is given by the XXZ $S=1$ Heisenberg model with single-ion
anisotropy:\begin{equation}
H=\sum_{i}\left[S_{i}^{x}S_{i+1}^{x}+S_{i}^{y}S_{i+1}^{y}+\lambda S_{i}^{z}S_{i+1}^{z}+D\left(S_{i}^{z}\right)^{2}\right].\label{eq:XXZ1}\end{equation}
This model exhibits a very rich ground state phase diagram (see \cite{chen03,cristian03}
and references therein) with six different phases. According to numerical
studies \cite{verstraete04b}, the LE is one at the isotropic Heisenberg
point. In the present work it is proved that the LE is always one
in the entire region of parameters, in spite of the complexity of
the phase diagram. Of course this makes unpractical the use of the
LE for the detection of quantum phase transition in contrast to what
suggested by some authors \cite{verstraete04b,popp04}. 

One may wonder whether the maximal localizability of the entanglement
is due to the presence of the two qbits at the borders. This occurs,
for example, in the large-$D$ region where the asymptotic ground
state is $\left(|\uparrow,0,\ldots,0,\downarrow\rangle-|\downarrow,0,\ldots,0,\uparrow\rangle\right)/\sqrt{2}$
for which the LE is one, while in absence of the spin 1/2 at the endpoints,
the ground state tends to the unentangled state $|0,\ldots,0\rangle$.

A little care is needed in the broken symmetry phases i.e.~ferro-
and antiferro-magnetic regions where the ground states are twofold
degenerate. However the requirement of invariance under $\Pi^{x,y,z}$
fixes the combination of ground states in a similar manner as in the
Greenberger-Horne-Zeilinger (GHZ) state with maximal LE.

\emph{Matrix Product States.} Matrix product states (MPS) are very
interesting in many respects, in particular the fixed point of the
density matrix renormalization group algorithm (DMRG) yields an MPS
state \cite{ostlund95}. As was shown in Ref.~\cite{verstraete04b},
it is possible to calculate exactly the LE for any MPS state. This
opens up the possibility of calculating approximately the LE for any
one dimensional ground state. Here we revisit the examples introduced
in \cite{verstraete04b} in the framework of our formalism. This will
allow us to clarify the link between LE and correlations.

Using the convention of \cite{verstraete04b} we write a general MPS
state with open boundary condition as\begin{equation}
|F_{A}\rangle=\mathcal{C}\sum_{\beta_{1},\cdots,\beta_{N}}|\beta_{1}\cdots\beta_{N}\rangle\left(\1\otimes A^{\beta_{N}}\cdots A^{\mathbf{\beta}_{1}}\right)|\Psi^{-}\rangle_{0,N+1},\label{eq:MPS}\end{equation}
where the $A^{\beta}$'s are $D$ $2\times2$ matrices which depend
on the local basis $|\beta_{i}\rangle$, $\mathcal{C}$ is a normalization
constant and $|\Psi^{-}\rangle_{0,N+1}=\left(|\uparrow\downarrow\rangle_{0,N+1}-|\downarrow\uparrow\rangle_{0,N+1}\right)/\sqrt{2}$
is the two-qbit singlet at the endpoints.

The optimal basis $|s_{i}\rangle$ can be found via the method shown
in \cite{verstraete04b} and, since in this case the preconcurrence
takes the form $-\det\left(A^{s_{N}}\cdots A^{s_{2}}A^{s_{1}}\right)$,
the $G$ operator factorizes into local terms reducing to \[
G^{\left(s\right)}=-\prod_{i=1}^{N}G_{i},\qquad G_{i}=\sum_{s_{i}}\textrm{sign}\left(\det\left(A^{s_{i}}\right)\right)|s_{i}\rangle\langle s_{i}^{\ast}|.\]
 Hence the localizable entanglement is\begin{equation}
L\left(|F\rangle\right)=-\langle F|\sigma_{0}^{y}G_{1}G_{2}\cdots G_{N}\sigma_{N+1}^{y}|F^{\ast}\rangle.\end{equation}

A well known example of MPS state is the ground state of the spin-1
AKLT model, characterized by the presence of a gap, exponentially
decaying spin-spin correlation function and hidden topological (string)
order \cite{AKLT}. As any MPS, the AKLT ground state can be alternatively
written in the valence bond solid (VBS) formalism. In this representation,
every spin one at position $i$ is replaced by a pair of qbits $i,\bar{i}$
each one forming a singlet with its nearest neighbor, respectively
$\overline{i-1}$ and $i+1$. The on-site couple of qbits is then
projected back onto the spin-1 Hilbert space by means of a single
$3\times4$ matrix $A$, related to the $2\times2$ MPS matrices $A^{\beta}$
\emph{}via $\langle\beta_{i}|A=\langle\Psi^{-}|_{i\bar{i}}A^{\beta_{i}}\otimes\1_{\bar{i}}$. 

The $\phi$-deformed generalization $|V_{\phi}\rangle$ \cite{verstraete04b}
of the AKLT state is given by the matrix \[
A=\left(\begin{array}{cccc}
e^{\phi} & 0 & 0 & 0\\
0 & \frac{e^{-\phi}}{\sqrt{2}} & \frac{e^{\phi}}{\sqrt{2}} & 0\\
0 & 0 & 0 & e^{-\phi}\end{array}\right).\]
The $|V_{\phi=0}\rangle$ state is the rotation-invariant AKLT state,
while a non zero $\phi$ breaks the $O\left(3\right)$ invariance
down to $O\left(2\right)$ invariance. As was shown in \cite{verstraete04b}
the optimal basis is $\left\{ |\tilde{s}\rangle\right\} =\left\{ |0\rangle,|\pm\rangle\right\} $
for any $\phi$, which is the same as in the previously considered
$S=1$ case. It is straightforward to calculate the $G$ operator
and the LE which now reads \begin{equation}
G^{\left(\tilde{s}\right)}\left(|V_{\phi}\rangle\right)=-\prod_{i=1}^{N}e^{i\pi S_{i}^{y}},\,\,\, L\left(|V_{\phi}\rangle\right)=\langle V_{\phi}|e^{i\pi S_{\textrm{tot}}^{y}}|V_{\phi}\rangle,\label{eq:LE-VBS}\end{equation}
 For $\phi=0$ the LE is equal to one, while for non zero $\phi$
it decays exponentially with a finite entanglement length $\xi_{E}^{-1}=-\lim_{N\rightarrow\infty}\ln\left(L\right)/N$.
On the basis of Eq. (\ref{eq:LE-VBS}), the connection between the
LE and the string order parameter (SOP) in the $y$ direction becomes
evident. Generally the (maximally extended) SOP's $g^{\alpha}\left(N\right)$
are defined as \[
g^{\alpha}\left(N\right)=-\langle V_{\phi}|S_{1}^{\alpha}\prod_{i=2}^{N-1}e^{i\pi S_{i}^{\alpha}}S_{N}^{\alpha}|V_{\phi}\rangle.\]
 In the isotropic case $\phi=0$, all $g^{x,y,z}$ saturate to $4/9$
revealing a breaking of a hidden $\mathbb{Z}_{2}\times\mathbb{Z}_{2}$
symmetry \cite{kennedy92}. On the other hand, for $\phi\neq0$ the
SOP in the $z$ direction tends exponentially to a non-zero value
$g^{z}\left(\infty\right)=4\left(\cosh\left(2\phi\right)+\sqrt{\cosh\left(2\phi\right)^{2}+3}\right)^{-2}$,
while $g^{y}$ (and $g^{x}$) behaves as the LE decreasing exponentially
to zero with correlation length $\xi_{E}$. In the $\phi=0$ case
the LE can be ascribed to the presence of string order. However in
the deformed case, it was argued in \cite{popp04} that the LE is
not connected to string order since the LE is short ranged while $g^{z}$
is non vanishing. Actually our arguments show that in this case a
connection still exists between the LE and SOP, but in the $y$ channel. 

An example of MPS state $|\eta\rangle$ for which the LE is identically
one but all the SOP's are zero was conceived in \cite{popp04} and
is defined as in (\ref{eq:MPS}) with $A^{1}=\sigma^{z}+\sigma^{y},$
$A^{2}=\sigma^{z}-i\1,$ and $\left\{ |\beta\rangle\right\} $ is
the standard basis which in this case coincides with the optimal one.
It is straightforward to calculate the $G$ operator for this state
which turns out to be $G=-\left(-1\right)^{N}\1$ yielding a maximal
LE: \begin{equation}
L\left(|\eta\rangle\right)=-\left(-1\right)^{N}\langle\eta|\sigma_{0}^{y}\sigma_{N+1}^{y}|\eta^{\ast}\rangle=1.\end{equation}
It is clear that, in this case, the LE is not connected to string
parameters but rather to a classical correlation that involves a complex
conjugation.

In conclusion, in this Letter we have considered the recently defined
LE and its relationship with correlations in spin system. In quantum
information theory the LE has a well defined physical meaning, while
in condensed matter it has attracted much attention for its possible
uses to detect and characterize quantum phase transitions. In particular,
we have shown that for spin-1 systems parity symmetry along the three
axes is sufficient to assure that the LE reaches its maximal value
one. This promotes spin-1 systems as realizations of perfect quantum
channels, but shows that the LE is insensitive to quantum phase transitions.
Using similar symmetry arguments we have put in evidence that, in
the spin-1/2 case, the LE equals the maximal correlation. Both of
these findings have been obtained analytically confirming the numerical
results of Ref.~\cite{popp04}. We believe that relaxing the symmetry
requirements and/or analyzing higher spin systems opens the possibility
of finding examples in which the LE does not reduce to known correlations. 

We are grateful to C. Degli Esposti Boschi, E. Ercolessi, G. Morandi,
F. Ortolani and S. Pasini for useful discussions. This work was partially
supported by the TMR network EUCLID (contract number: HPRN-CT-2002-00325),
and the italian MIUR through COFIN projects (prot.~n.~2002024522\_001
and 2003029498\_013).  

\bibliographystyle{apsrev}
\bibliography{/home/campos/myart/final/resubmission/entang}

\begin{thebibliography}{17}
\expandafter\ifx\csname natexlab\endcsname\relax\def\natexlab#1{#1}\fi
\expandafter\ifx\csname bibnamefont\endcsname\relax
  \def\bibnamefont#1{#1}\fi
\expandafter\ifx\csname bibfnamefont\endcsname\relax
  \def\bibfnamefont#1{#1}\fi
\expandafter\ifx\csname citenamefont\endcsname\relax
  \def\citenamefont#1{#1}\fi
\expandafter\ifx\csname url\endcsname\relax
  \def\url#1{\texttt{#1}}\fi
\expandafter\ifx\csname urlprefix\endcsname\relax\def\urlprefix{URL }\fi
\providecommand{\bibinfo}[2]{#2}
\providecommand{\eprint}[2][]{\url{#2}}

\bibitem[{\citenamefont{Schrödinger}(1935)}]{schroedinger35}
\bibinfo{author}{\bibfnamefont{E.}~\bibnamefont{Schrödinger}},
  \bibinfo{journal}{Proc. Cambridge Philos. Soc.}
  \textbf{\bibinfo{volume}{31}}, \bibinfo{pages}{555} (\bibinfo{year}{1935}).

\bibitem[{\citenamefont{Heiss}(2002)}]{heissbook}
\bibinfo{editor}{\bibfnamefont{D.}~\bibnamefont{Heiss}}, ed.,
  \emph{\bibinfo{title}{Fundamentals of Quantum Information}}
  (\bibinfo{publisher}{Springer}, \bibinfo{year}{2002}).

\bibitem[{\citenamefont{Osborne and Nielsen}(2002)}]{osborne02}
\bibinfo{author}{\bibfnamefont{T.~J.} \bibnamefont{Osborne}} \bibnamefont{and}
  \bibinfo{author}{\bibfnamefont{M.~A.} \bibnamefont{Nielsen}},
  \bibinfo{journal}{Phys.~Rev.~A} \textbf{\bibinfo{volume}{66}},
  \bibinfo{pages}{32110} (\bibinfo{year}{2002}).

\bibitem[{\citenamefont{Vidal et~al.}(2003)\citenamefont{Vidal, Latorre, Rico,
  and Kitaev}}]{vidal03}
\bibinfo{author}{\bibfnamefont{G.}~\bibnamefont{Vidal}},
  \bibinfo{author}{\bibfnamefont{J.~I.} \bibnamefont{Latorre}},
  \bibinfo{author}{\bibfnamefont{E.}~\bibnamefont{Rico}}, \bibnamefont{and}
  \bibinfo{author}{\bibfnamefont{A.}~\bibnamefont{Kitaev}},
  \bibinfo{journal}{\PRL} \textbf{\bibinfo{volume}{90}},
  \bibinfo{pages}{227902} (\bibinfo{year}{2003}).

\bibitem[{\citenamefont{Fan et~al.}(2004)\citenamefont{Fan, Korepin, and
  Roychowdhury}}]{fan04}
\bibinfo{author}{\bibfnamefont{H.}~\bibnamefont{Fan}},
  \bibinfo{author}{\bibfnamefont{V.}~\bibnamefont{Korepin}}, \bibnamefont{and}
  \bibinfo{author}{\bibfnamefont{V.}~\bibnamefont{Roychowdhury}},
  \bibinfo{journal}{\PRL} \textbf{\bibinfo{volume}{93}},
  \bibinfo{pages}{227203} (\bibinfo{year}{2004}).

\bibitem[{\citenamefont{Verstraete
  et~al.}(2004{\natexlab{a}})\citenamefont{Verstraete, Popp, and
  Cirac}}]{verstraete04a}
\bibinfo{author}{\bibfnamefont{F.}~\bibnamefont{Verstraete}},
  \bibinfo{author}{\bibfnamefont{M.}~\bibnamefont{Popp}}, \bibnamefont{and}
  \bibinfo{author}{\bibfnamefont{J.~I.} \bibnamefont{Cirac}},
  \bibinfo{journal}{\PRL} \textbf{\bibinfo{volume}{92}},
  \bibinfo{pages}{027901} (\bibinfo{year}{2004}{\natexlab{a}}).

\bibitem[{\citenamefont{Briegel et~al.}(1998)\citenamefont{Briegel, D\"{u}r,
  Cirac, and Zoller}}]{briegel98}
\bibinfo{author}{\bibfnamefont{H.-J.} \bibnamefont{Briegel}},
  \bibinfo{author}{\bibfnamefont{W.}~\bibnamefont{D\"{u}r}},
  \bibinfo{author}{\bibfnamefont{J.~I.} \bibnamefont{Cirac}}, \bibnamefont{and}
  \bibinfo{author}{\bibfnamefont{P.}~\bibnamefont{Zoller}},
  \bibinfo{journal}{\PRL} \textbf{\bibinfo{volume}{81}}, \bibinfo{pages}{5932}
  (\bibinfo{year}{1998}).

\bibitem[{\citenamefont{Ursin et~al.}(2004)\citenamefont{Ursin, Jennewein,
  Aspelmeyer, Kaltenbaek, Lindenthal, Walther, and
  Zeilinger}}]{zeilinger-danube}
\bibinfo{author}{\bibfnamefont{R.}~\bibnamefont{Ursin}},
  \bibinfo{author}{\bibfnamefont{T.}~\bibnamefont{Jennewein}},
  \bibinfo{author}{\bibfnamefont{M.}~\bibnamefont{Aspelmeyer}},
  \bibinfo{author}{\bibfnamefont{R.}~\bibnamefont{Kaltenbaek}},
  \bibinfo{author}{\bibfnamefont{M.}~\bibnamefont{Lindenthal}},
  \bibinfo{author}{\bibfnamefont{P.}~\bibnamefont{Walther}}, \bibnamefont{and}
  \bibinfo{author}{\bibfnamefont{A.}~\bibnamefont{Zeilinger}},
  \bibinfo{journal}{Nature (London)} \textbf{\bibinfo{volume}{430}},
  \bibinfo{pages}{849} (\bibinfo{year}{2004}).

\bibitem[{\citenamefont{Wootters}(1998)}]{wootters98}
\bibinfo{author}{\bibfnamefont{W.~K.} \bibnamefont{Wootters}},
  \bibinfo{journal}{\PRL} \textbf{\bibinfo{volume}{80}}, \bibinfo{pages}{2245}
  (\bibinfo{year}{1998}).

\bibitem[{\citenamefont{DiVincenzo et~al.}(1998)\citenamefont{DiVincenzo,
  Fuchs, Mabuchi, Smolin, Thapliyal, and Uhlmann}}]{divincenzo98}
\bibinfo{author}{\bibfnamefont{D.~P.} \bibnamefont{DiVincenzo}},
  \bibinfo{author}{\bibfnamefont{C.~A.} \bibnamefont{Fuchs}},
  \bibinfo{author}{\bibfnamefont{H.}~\bibnamefont{Mabuchi}},
  \bibinfo{author}{\bibfnamefont{J.~A.} \bibnamefont{Smolin}},
  \bibinfo{author}{\bibfnamefont{A.}~\bibnamefont{Thapliyal}},
  \bibnamefont{and} \bibinfo{author}{\bibfnamefont{A.}~\bibnamefont{Uhlmann}},
  \bibinfo{journal}{\mbox{quant-ph/9803033}}  (\bibinfo{year}{1998}).

\bibitem[{\citenamefont{Popp et~al.}(2004)\citenamefont{Popp, Verstraete,
  Mart\'{\i}n-Delgado, and Cirac}}]{popp04}
\bibinfo{author}{\bibfnamefont{M.}~\bibnamefont{Popp}},
  \bibinfo{author}{\bibfnamefont{F.}~\bibnamefont{Verstraete}},
  \bibinfo{author}{\bibfnamefont{M.~A.} \bibnamefont{Mart\'{\i}n-Delgado}},
  \bibnamefont{and} \bibinfo{author}{\bibfnamefont{J.}~\bibnamefont{Cirac}},
  \bibinfo{journal}{quant-ph/0411123}  (\bibinfo{year}{2004}).

\bibitem[{\citenamefont{Verstraete
  et~al.}(2004{\natexlab{b}})\citenamefont{Verstraete, Mart\'{\i}n-Delgado, and
  Cirac}}]{verstraete04b}
\bibinfo{author}{\bibfnamefont{F.}~\bibnamefont{Verstraete}},
  \bibinfo{author}{\bibfnamefont{M.~A.} \bibnamefont{Mart\'{\i}n-Delgado}},
  \bibnamefont{and} \bibinfo{author}{\bibfnamefont{J.~I.} \bibnamefont{Cirac}},
  \bibinfo{journal}{\PRL} \textbf{\bibinfo{volume}{92}},
  \bibinfo{pages}{087201} (\bibinfo{year}{2004}{\natexlab{b}}).

\bibitem[{\citenamefont{Affleck et~al.}(1988)\citenamefont{Affleck, Kennedy,
  Lieb, and Tasaki}}]{AKLT}
\bibinfo{author}{\bibfnamefont{I.}~\bibnamefont{Affleck}},
  \bibinfo{author}{\bibfnamefont{T.}~\bibnamefont{Kennedy}},
  \bibinfo{author}{\bibfnamefont{E.~H.} \bibnamefont{Lieb}}, \bibnamefont{and}
  \bibinfo{author}{\bibfnamefont{H.}~\bibnamefont{Tasaki}},
  \bibinfo{journal}{Commun.~Math.~Phys.} \textbf{\bibinfo{volume}{115}},
  \bibinfo{pages}{477} (\bibinfo{year}{1988}).

\bibitem[{\citenamefont{Chen et~al.}(2003)\citenamefont{Chen, Hida, and
  Sanctuary}}]{chen03}
\bibinfo{author}{\bibfnamefont{W.}~\bibnamefont{Chen}},
  \bibinfo{author}{\bibfnamefont{K.}~\bibnamefont{Hida}}, \bibnamefont{and}
  \bibinfo{author}{\bibfnamefont{B.~C.} \bibnamefont{Sanctuary}},
  \bibinfo{journal}{\PRB} \textbf{\bibinfo{volume}{67}},
  \bibinfo{pages}{104401} (\bibinfo{year}{2003}).

\bibitem[{\citenamefont{\mbox{Degli Esposti Boschi}
  et~al.}(2003)\citenamefont{\mbox{Degli Esposti Boschi}, Ercolessi, Ortolani,
  and Roncaglia}}]{cristian03}
\bibinfo{author}{\bibfnamefont{C.}~\bibnamefont{\mbox{Degli Esposti Boschi}}},
  \bibinfo{author}{\bibfnamefont{E.}~\bibnamefont{Ercolessi}},
  \bibinfo{author}{\bibfnamefont{F.}~\bibnamefont{Ortolani}}, \bibnamefont{and}
  \bibinfo{author}{\bibfnamefont{M.}~\bibnamefont{Roncaglia}},
  \bibinfo{journal}{Eur. Phys. J. B} \textbf{\bibinfo{volume}{35}},
  \bibinfo{pages}{465} (\bibinfo{year}{2003}).

\bibitem[{\citenamefont{\"{O}stlund and Rommer}(1995)}]{ostlund95}
\bibinfo{author}{\bibfnamefont{S.}~\bibnamefont{\"{O}stlund}} \bibnamefont{and}
  \bibinfo{author}{\bibfnamefont{S.}~\bibnamefont{Rommer}},
  \bibinfo{journal}{\PRL} \textbf{\bibinfo{volume}{75}}, \bibinfo{pages}{3537}
  (\bibinfo{year}{1995}).

\bibitem[{\citenamefont{Kennedy and Tasaki}(1992)}]{kennedy92}
\bibinfo{author}{\bibfnamefont{T.}~\bibnamefont{Kennedy}} \bibnamefont{and}
  \bibinfo{author}{\bibfnamefont{H.}~\bibnamefont{Tasaki}},
  \bibinfo{journal}{Commun. Math. Phys.} \textbf{\bibinfo{volume}{147}},
  \bibinfo{pages}{431} (\bibinfo{year}{1992}).

\end{thebibliography}

\end{document}